\documentclass[cits]{PoS}

\title{Measurement of the dark matter velocity anisotropy profile in galaxy clusters}
\ShortTitle{Dark matter velocity anisotropy}

\author{\speaker{Ole Host}\\
Dark Cosmology Centre, Niels Bohr Institute, University of Copenhagen, Juliane Maries Vej 30, DK-2100 Copenhagen, Denmark\\
E-mail: \email{olehost@dark-cosmology.dk}}

\abstract{Dark matter particles form halos that contribute the major part of the mass of galaxy clusters. The formation of these cosmological structures have been investigated both observationally and in numerical simulations, which have confirmed the existence of a universal mass profile \cite{1996ApJ...462..563N}. However, the dynamic behaviour of dark matter in halos is not as well understood. We have used observations of 16 equilibrated galaxy clusters to show that the random velocities of dark matter particles are larger on average along the radial direction than along the tangential, and that the magnitude of this velocity anisotropy is radially varying \cite{2008arXiv0808.2049H}. Our measurement implies that the collective behaviour of dark matter particles is fundamentally different from that of normal particles and the radial variation of the anisotropy velocity agrees with the predictions of numerical simulation \cite{1996MNRAS.281..716C}.}

\FullConference{Identification of dark matter 2008\\
		 August 18-22, 2008\\
		 Stockholm, Sweden}

\begin{document}

Dark matter is most often assumed to be a new type of particle which couples only very weakly to itself or other particle species \cite{2005PhR...405..279B}. Unlike a collisional gas, dark matter structures may therefore have a non-zero velocity anisotropy defined as $\beta = 1 - \sigma^2_t/\sigma^2_r$,
where $\sigma^2_t$ and $\sigma^2_r$ are the 1-dimensional tangential and radial velocity dispersions of the dark matter. Here we summarize the measurement of the radial velocity anisotropy profile in a sample of galaxy clusters, as presented in \cite{2008arXiv0808.2049H}.

Most of the ordinary baryonic matter in galaxy clusters is found in the intracluster medium (ICM), a hot plasma emitting X-rays. We consider a sample of 16 relaxed galaxy clusters for which the radial ICM density and temperature profiles have been obtained from X-ray observations in earlier work. The clusters are selected to appear close to circular in projection, have smooth gas temperature and density profiles, and reconstructed total density profiles which are monotonically declining. The sample consists of eleven highly relaxed cool-core clusters at low redshift observed with {\it XMM-Newton} (A262, A496, A1795, A1837, A2052, A4059, S\'ersic 159$-$3, MKW3s, MKW9, NGC533, and 2A0335+096) \cite{2004A&A...413..415K,2005A&A...433..101P}, and five intermediate redshift clusters observed with {\it Chandra} (RXJ1347.5, A1689, A2218, A1914, and A611) \cite{Morandi:2007aw}. Data for A2052 and S\'ersic 159$-$3 were also used in an earlier analysis \cite{2007A&A...476L..37H} where a constant velocity anisotropy was assumed. 

\begin{figure}
\begin{center}
\includegraphics[width=.7\textwidth]{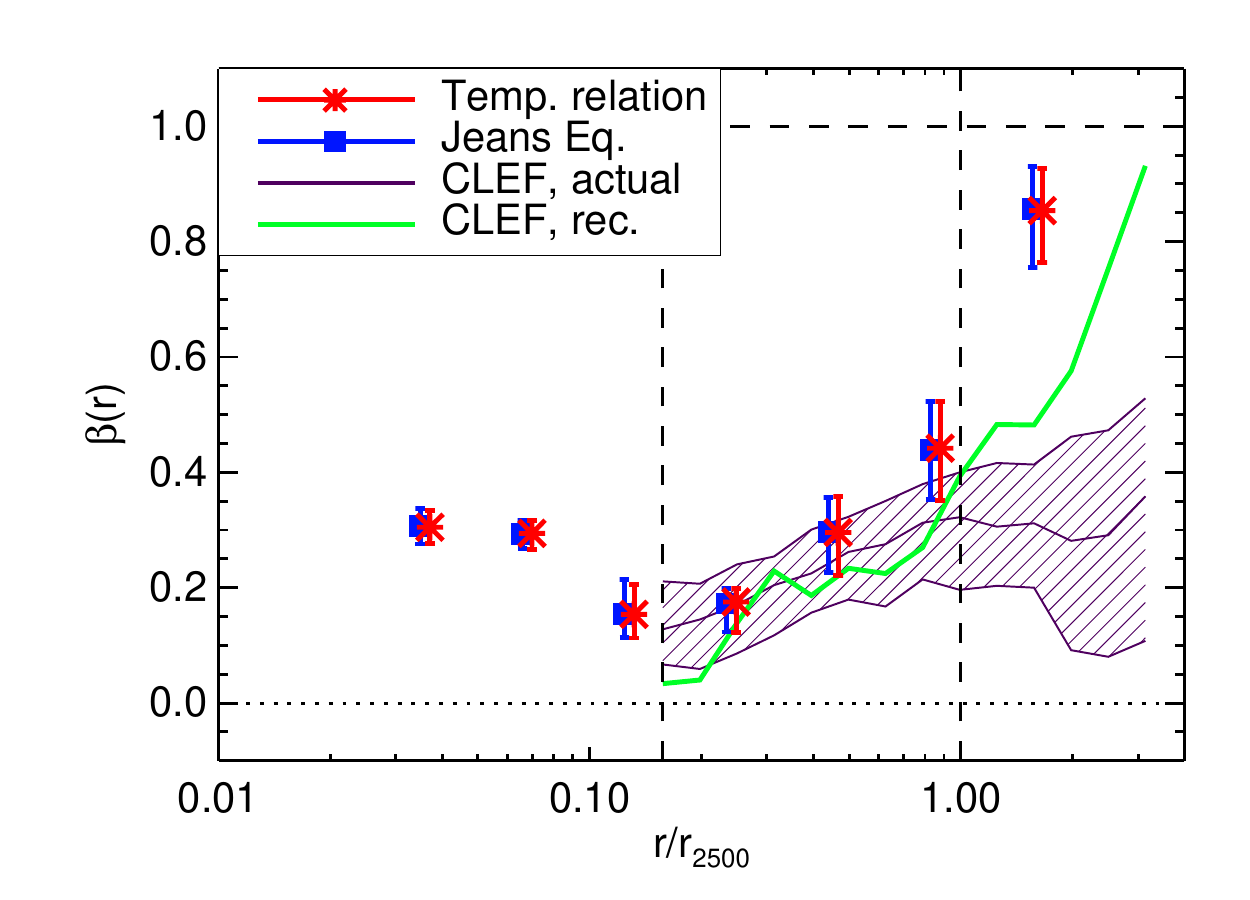}
\caption{Radial profile of the stacked velocity anisotropy for our sample of 16 clusters. There are two slightly different ways of determining the velocity anisotropy. The error bars indicate the $1\sigma$ percentiles of the underlying probability density. The hatched band shows the anisotropy profiles in the CLEF numerical simulation sample (see text), and the green line is the median anisotropy profile obtained when we apply our analysis to the CLEF sample. The correspondence between the reconstructed and actual velocity dispersions in the CLEF sample confirms the validity of our analysis. The leftmost vertical line indicates the innermost radius where the CLEF simulation resolves gravitational forces reliably, while the rightmost vertical line marks the onset of significant deviations from hydrostatic support in the CLEF sample. The horizontal dot-dashed line indicates the zero anisotropy of a gas.}\label{fi:bs}
\end{center}
\end{figure}

We determine the velocity anisotropy for the 16 clusters of our sample using the method outlined in \cite{2008arXiv0808.2049H}, and in almost all cases the anisotropy profile is radially increasing from close to zero in the center to about $0.5$ or greater in the outer parts. Since the qualitative behaviour of the profiles are similar, we combine all our data into a single `stacked' profile depending on radius in units of $r_{2500}$. Figure \ref{fi:bs} shows the resulting velocity anisotropy profile which varies between 0.1 and 0.5 in the radial range 0.03--1.0$\,r_{2500}$, where $r_{2500}$ is the scale radius enclosing a mean density 2500 times greater than the critical density at the redshift of the cluster. Our measurement is in good agreement with the predictions of numerical simulations (see below) in the overlapping radial range. However, at large radii $>r_{2500}$, we overestimate the velocity anisotropy due to a lack of hydrostatic support, as demonstrated in \cite{2008arXiv0808.2049H}.

\begin{figure}
\begin{center}
\includegraphics[width=.7\textwidth]{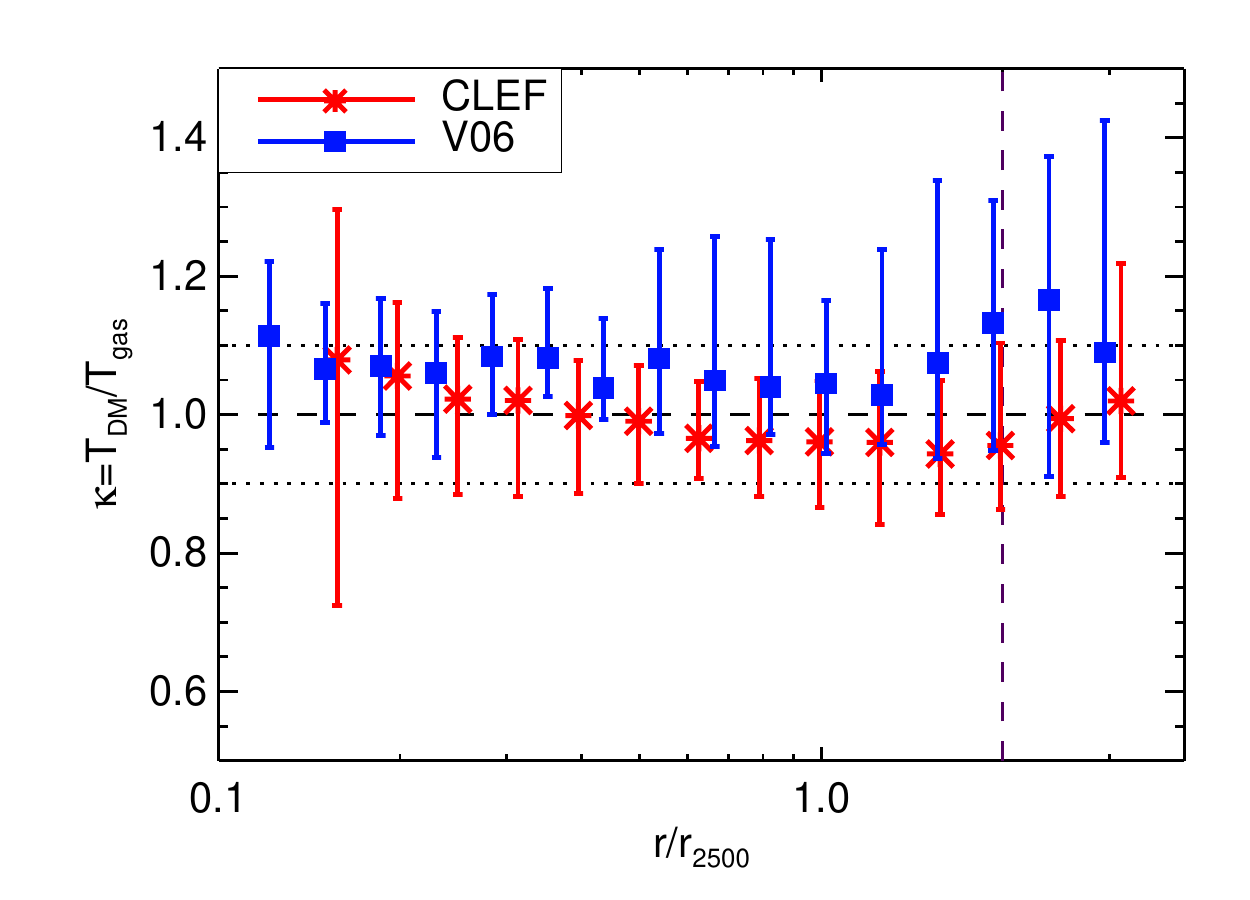}
\caption{Radial profile of $\kappa$ for the numerical simulation samples. The ratio of the specific energies is $\kappa=\sigma_\mathrm{DM}^2/\left(\frac{3k_\mathrm{B}T_{\mathrm{DM}}}{\mu m_H}\right),$ where $\mu=0.6$ is the mean molecular weight of the ICM gas particles. The vertical line indicates the largest radius of the observational data sample.}
\label{fig:kap}
\end{center}
\end{figure}

In order to calculate the velocity anisotropy, it is necessary to constrain the energy content of the dark matter. For this purpose we have assumed that the dark matter specific energy is equal to that of the ICM gas at the same radius up to a factor $\kappa$, of order unity. This assumption is validated by the results of two independent numerical simulations of the formation of galaxy clusters in the $\Lambda$CDM cosmology, which both include dark matter and radiative gas physics. From each simulation we select the most relaxed clusters at redshift $z=0$, which yields two samples of 67 and 20 clusters that we refer to as the CLEF \cite{Kay:2006iz} and V06 \cite{2003MNRAS.339.1117V,2006NewA...12...71V,2008arXiv0808.1111P} samples. For these two samples we calculate the two median $\kappa$-profiles (Fig.~\ref{fig:kap}) and we find that in both cases $\kappa$ is close to unity in the full radial range which is resolved, with about 10\% scatter. Therefore, in our analysis we take $\kappa=1.0\pm0.1$. With this $\kappa$-profile, it is possible to calculate the radial velocity dispersion and the velocity anisotropy of the dark matter. This is done strictly using non-parametric Monte Carlo methods so that no modeling of the ICM or dark matter properties is involved. 

We have made a number of tests where we assume different $\kappa$-profiles, including both profiles that are greater or lesser than unity, and radially varying profiles which mimick the most extreme radial variations compatible with the CLEF and V06 $\kappa$-profiles. In particular, we check the effect of both AGN heating ($\kappa<1$) and radiative cooling ($\kappa>1$) in the center of the clusters where simulations do not probe. In all cases the general behaviour is the same, i.e.~a radially increasing velocity anisotropy, although in some cases ($\kappa<1$) the velocity anisotropy can become zero at intermediate radii. See \cite{2008arXiv0808.2049H} for further details on these tests.

Our result shows that the collective behaviour of dark matter particles is fundamentally different from baryonic particles. Dark matter behaves as collisionless particles on the timescale of Gigayears, the dynamical timescale of galaxy clusters, and this implies an order--of--magnitude upper limit to the self-interaction cross-section per unit mass of $\sigma /m \lesssim 1\,$cm$^2$g$^{-1}$, similar to what was found for the Bullet Cluster \cite{2004ApJ...606..819M} and only slightly larger than the value proposed for self-interacting dark matter \cite{Spergel:1999mh}. According to numerical simulations, the Galactic dark matter halo is also anisotropic, and this anisotropy will influence direct detection rates at the 5-10\% level \cite{2008PhRvD..77b3509V} and is potentially measurable in directional searches \cite{2007JCAP...06..016H}.

\acknowledgments{The Dark Cosmology Centre is funded by the DNRF.}

\bibliographystyle{JHEP}
\bibliography{betacluster}

\end{document}